\documentclass[12pt]{article}
\usepackage{amsmath,amssymb,amsthm}
\usepackage{graphics,epsfig}
\usepackage{hyperref}
\usepackage{natbib}
\usepackage{color}
\usepackage{graphicx}
\usepackage{caption}
\usepackage{subcaption}
\usepackage{float}
\usepackage{mathrsfs}
\usepackage{multirow}
\usepackage{comment}
\usepackage{setspace}
\usepackage{bbm}
\usepackage{bm}
\usepackage{amsfonts}
\usepackage{xcolor}
\usepackage{pslatex}
\usepackage{setspace}


\begin{document}

  \title{\bfseries Strategic Complemantarities as stochastic control under sticky price}

  \author{
Lambert Dong\footnote{Corresponding author, {\small\texttt{lbd2321@jagmail.southalabama.edu}}}\; \footnote{Department of Mathematics and Statistics,  University of South Alabama, Mobile, AL, 36608,
United States.}
}

\date{\today}
\maketitle

\begin{abstract}
We examine how monetary shocks spread throughout an economic model characterized by sticky prices and general equilibrium, where the pricing strategies of firms are interlinked, fostering a mutually beneficial relationship. In this dynamic equilibrium, pricing choices of firms are influenced by overall economic factors, which are themselves affected by these decisions. We approach this situation using a path integral control method, yielding several important insights. We confirm the presence and uniqueness of the equilibrium and scrutinize the impulse response function (IRF) of output subsequent to a shock affecting the entire economy. We then inquire whether strategic interdependencies amplify the IRF over various time spans, potentially leading to a distinctive hump-shaped pattern. Additionally, we explore the correlation between the extent of interdependence and the IRF. If these factors exhibit an inverse relationship, we infer that the equilibrium collapses at a critical juncture. Notably, we validate these findings using both the Calvo and Golosov–Lucas models.	
\end{abstract}

\section{Introduction:}

Despite notable progress in both the theoretical and empirical realms of general equilibrium models featuring sticky prices, the imperative of tractability frequently leads to a neglect of the interplay among firms' pricing decisions \citep{pramanik2021consensus}. However, these complementarities are compelling owing to their potential to magnify the non-neutrality of nominal shocks, a notion advanced by \cite{nakamura2018high} and \cite{klenow2016real}, and their empirical relevance. Current general equilibrium analyses typically tackle these phenomena through numerical exploration, as evidenced in the works of \cite{nakamura2018high} and \cite{klenow2016real}, and \cite{mongey2021market}, or simplifying the consideration of timing adjustments, as in \cite{wang2022dynamic}, or excluding idiosyncratic shocks, as in \cite{caplin1991state}. We present an analytical framework for examining a general equilibrium where the dynamic evolution of aggregates influences individual decision-making, and vice versa \citep{pramanik2021optimization,pramanik2021optimal,pramanik2023scoring}. Our findings contribute to a comprehensive understanding of sticky-price equilibrium within a state-dependent model that incorporates both idiosyncratic shocks and strategic complementarities in pricing decisions. This methodological approach holds promise for applications across various domains within macroeconomics \citep{alvarez2023price}.

Examining strategic complementarities within a broader equilibrium framework presents intricate challenges: pivotal choices are contingent upon overarching factors influenced by individual decisions \citep{pramanik2020motivation,pramanik2023scoring}. This complexity gives rise to a fixed-point problem, particularly in models featuring abrupt shifts, wherein optimal choices exhibit nonlinearity and temporal evolution \citep{pramanik2022lock,pramanik2022stochastic,pramanik2023path1,pramanik2023path}.\cite{wang2022dynamic} discusses an analytic solutions of a dynamic oligopoly model \citep{hua2019assessing,pramanik2016tail,pramanik2021effects}. Their work achieves manageability by assuming externally determined timing for firm price adjustments, reminiscent of the Calvo model \citep{calvo1983staggered}. Our approach, akin to \cite{caplin1991state} and \cite{wang2022dynamic}, seeks to determine analytical solutions. However, a key difference lies in investigating a scenario where firm determinations are contingent on the prevailing conditions and where individualized shocks wield substantial influence at the firm level.

We present analytical finding that illuminates the optimal strategies of firms and the overall equilibrium dynamics within a dynamic framework characterized by strategic complementarities and state-dependent decisions.\cite{alvarez2023price} made a notable advancement by framing the problem within the mathematical framework of Mean Field Games (MFG)introduced by \cite{lasry2007mean}. This formulation yields a system of two interconnected partial differential equations: a backward Hamiltonian-Jacobi-Bellman (HJB) equation capturing individual decisions and a forward Fokker-Plank equation describing aggregation. The effectiveness of applying the MFG framework to analyze the dynamic behavior of high-dimensional cross-sections has been emphasized by \cite{achdou2022income}, as well as \cite{ahn2018inequality}, where numerical methods were discussed. Thus far, the development of MFG theory has primarily focused on two key aspects: firstly, scrutinizing the HJB equation and presenting an analytical solution in scenarios involving an infinite number of identical players; and secondly, examining differential games with a considerably large but finite number of identical players, and establishing connections between their asymptotic tendencies as the number of players approaches infinity and the HJB equation. The first aspect is thoroughly comprehended and extensively documented. As for the second aspect, the asymptotic behavior of differential games as the number of players approaches infinity has been comprehended in the context of ergodic differential games. However, the broader applicability of this understanding remains an ongoing inquiry \citep{cardaliaguet2010notes}.

\subsection{Main contribution:}
We intend to employ the Feynman-type path integral method developed by \cite{pramanik2020optimization} and further elaborated upon by \cite{pramanik2024optimization} to derive an analytical solution for the aforementioned system \citep{pramanik2023semicooperation}. When the state variable possesses a considerably large dimension and the system dynamics exhibit nonlinearity, such as in Merton-Garman-Hamiltonian stochastic differential equations (SDEs), constructing an HJB equation numerically becomes exceedingly challenging. The Feynman-type path integral approach effectively addresses this dimensionality obstacle and furnishes a localized analytic solution. To apply this approach, we initially formulate a stochastic Lagrangian for individual continuous time points within the interval $s\in[0,t]$ where $t>0$. Subsequently, we partition this entire time span into $n$-number of equal-length subintervals and establish a Riemann measure corresponding to the state variable for each subinterval. Following the construction of a Euclidean action function, we derive a Schr\"odinger-type equation through Wick rotation. By enforcing the first-order conditions with respect to both the state and control variables, we ascertain the solution of the system. This approach can be applicable in cancer research \citep{dasgupta2023frequent,hertweck2023clinicopathological,kakkat2023cardiovascular,khan2023myb,vikramdeo2023profiling}.

We establish criteria for the existence and uniqueness of equilibrium discussed above and offer a comprehensive analytical examination of the IRF of output following a one-time nominal shock. Furthermore, we examine whether the intensification of strategic complementarity leads to a convex increment in the IRF of output at each stage, signifying a more notable rise with increased levels of complementarity. Under this circumstance, the MFG approach suggests that the IRF approaches infinity as strategic complementarity approaches a critical threshold. Beyond this threshold, equilibrium dissipates, and for values surpassing it, equilibrium lacks stability, potentially exhibiting discontinuities in parameter variations. Around these critical values where equilibrium breaks down, there are significant and sudden shifts in equilibrium outcomes, known as \emph{pole} effects. Conversely, when interactions demonstrate substitutability instead of complementarity, equilibrium always persists, and the IRF converges to zero as substitutability tends toward infinity, mirroring an economy with flexible prices. The absence of multiple equilibria, despite substantial strategic complementarity, may appear unexpected. We also perform some numerical simulation of our results and do a real data analysis of this model.

\section{Background literature:}

Our integration of strategic complementarities resonates with the influential work of \cite{caplin1991state}, where the profit function of a firm hinges on both its individual markup and the average markup. One notable distinction is the incorporation of idiosyncratic shocks into our economic framework, a component absents from their analysis. While Caplin and Leahy investigated an equilibrium scenario characterized by aggregate nominal shocks following a drift-less Brownian motion, our primary emphasis revolves around examining the impulse response subsequent to a one-time shock with a non-zero drift. This type of construction gives us more freedom to study a generalized system.

Our research closely aligns with the studies conducted by \cite{nakamura2018high} and \cite{klenow2016real}. In both studies, Dynamic Stochastic General Equilibrium (DSGE) models incorporate an input-output structure, where the sticky price of other industries factors into the cost for each industry, showcasing \emph{macro strategic complementarities}. Similarly, all three studies account for a frictionless labor market, idiosyncratic shocks at the firm level, and menu costs incurred by firms for price adjustments. Nakamura and Steinsson (2010)\citep{nakamura2018high}, like our approach, allow for random menu costs, while Klenow and Willis (2016)\citep{klenow2016real}, also similar to our approach, incorporate non-constant demand elasticity at the firm level, termed \emph{micro-strategic complementarities}. We demonstrate that, up to the second order, the combined effect of micro and macro complementarities can be represented through a single parameter. While both \cite{nakamura2018high} and \cite{klenow2016real} employ numerical techniques to investigate the impact of monetary shocks on aggregate output, our study presents analytical findings.

Our analysis is also pertinent to the research conducted by \cite{wang2022dynamic}, who analyses the propagation of shocks within a sticky-price economy marked by strategic complementarities. They provide an analytical solution assuming that firms adhere to a time-dependent rule akin to the Calvo framework. Certain aspects of the underlying environment parallel our study: the factors contributing to complementarities, such as variable demand elasticity, diminishing returns, and non-zero Frisch elasticity, are captured by a single parameter. However, there are differences in modeling approaches: firstly, they explore a dynamic oligopoly scenario without idiosyncratic shocks, whereas our focus lies on oligopolistically competitive markets featuring idiosyncratic shocks—a characteristic advantageous for aligning with observed price change distributions in real-world data. Secondly, in their analysis, the timing of adjustments is predetermined, whereas in our framework, firms determine both the timing and magnitude of price adjustments. The simplification of exogenous timing and the absence of idiosyncratic shocks in their model facilitate connections with the New Keynesian Phillips curve and permit an investigation into the importance of strategic complementarities. 

\cite{bertucci2018optimal} contributes to the MFG literature by investigating a problem concerning impulse control. However, his analysis is simpler, involving a decision maker contemplating only one adjustment, with the adjustment target predetermined. Additionally, his emphasis is on establishing the existence and uniqueness of solutions, employing a slightly different notion of solution.

\section{Basic Calvo model:}
This section explores an issue concerning strategic complementarities and pricing, as originally discussed by\cite{calvo1983staggered}. This particular scenario garners frequent attention in studies on sticky prices owing to its practical significance. The model offers a clear-cut framework for elucidating the core elements of the analysis and for scrutinizing crucial outcomes such as the existence, uniqueness, and non-monotonic characteristics of impulse response profiles, which also bear relevance to the state-dependent problem \citep{pramanik2024bayes}.

Let at time $s$, $Z(s)$ be the consumer price index (CPI), $V_i(s)$ be a consumer preference shock corresponding to  $i^{th}$ variety and the price set by a firm on consumer good of $i^{th}$ variety be $\hat Z(s)$ such that $z(s)\equiv\hat z(s)/V_i(s)$. Define $\tilde x(s):=\frac{ z(s)-\hat Z(s)}{Z(s)}$ and $\tilde X(s):=\frac{Z(s)-\hat Z(s)}{Z(s)}$ as percent deviation from symmetric equilibrium of a firm's own and the aggregate price (CPI), respectively. The economy comprises a range of \emph{atomistic} individual firms, each operating independently. Each firm operates under the assumption of a consistent fluctuation in markup averages, denoted by $\tilde X(s)$ for all times $s\in[0,t]$. The firm has the ability to adjust its pricing only at specific, randomly occurring times denoted by $\{\xi_i\}$, which follows a Poisson process characterized by a parameter $\theta$. These instances of adjustment are referred to as \emph{adjustment opportunities}, and the state of the firm's pricing at these times is termed the \emph{optimal reset value}. Following a price adjustment at time s, the difference in markup, $\tilde x(s)$, jumps according to a Brownian motion without a drift component but with a variance of $\sigma^2$. Additionally, the markup experiences abrupt jumps immediately after a price adjustment at $s=\xi_i$, with each jump in markup denoted by $\mathcal U_i$. Therefore, the markup gap evolves as
\begin{equation}\label{1}
\tilde X(s)=\tilde X(0)+\sigma\left[{\bf \mathcal W}(s)-{\bf \mathcal W}(0)\right]+\sum_{\xi_i\leq s} \mathcal \mathcal U_i,\ \ \text{for all $s\in[0,t]$}, 
\end{equation}
where $\bf \mathcal W$ is a m-dimensional standard Brownian motion. Furthermore, in the absence of any markup jump the continuous version of Equation (\ref{1}) becomes,
\[
\tilde X(t)=\tilde X(0)+\int_0^t\sigma d\mathcal W(s).
\]
Throughout our analysis we use the following form of the SDE
\[
d\tilde X(s)=\mu[s,u(s),\tilde X(s)]ds+\sigma[s,u(s),\tilde X(s)]d\mathcal W(s),
\]
where $\mu[.]$ and $\sigma[.]$ are the drift and diffusion components, respectively and $u(s)$ is the complementarity strategy of a firm at time $s$. The above SDE would be the constraint of our whole analysis.

\section{Broader Impact:}
We investigate the propagation of monetary shocks within a sticky-price general equilibrium model, wherein the decisions of a firm's price involve strategic interactions with other firms' decisions. This problem is intricate, and a comprehensive analytic characterization of the determinants of equilibrium dynamics is lacking \citep{polansky2021motif,pramanik2024estimation}. We reformulate the fixed-point problem defining equilibrium as a path integral control and derive several analytical findings concerning equilibrium existence and the analytic characterization of an impulse response \citep{pramanik2023optimal}.

The framework we examine is significant for examining equilibrium dynamics in related literature. \cite{alvarez2021empirical} utilized the equilibrium characterization established in Alvarez, Lippi, and Souganidis (2023)\citep{alvarez2023price} to analyze the impulse response to shocks featuring a transitory component, contrasting with the one-time shocks typically studied in literature. Furthermore, we extend this idea to investigate higher-order perturbations. This extension should be pivotal in comparing time- and state-dependent models, as these models respond differently to large versus small shocks.

	\bibliographystyle{apalike}
	\bibliography{bib}
\end{document}